\documentclass[prx,twocolumn,superscriptaddress]{revtex4}
\usepackage{graphicx,color}
\usepackage{amssymb}   
\usepackage{amsmath}
\usepackage{epstopdf}
\usepackage{natbib}
\usepackage{hyperref}
\usepackage{bm}
\usepackage{color}

\begin{document}
\title{Magnetic Field Driven Nodal Topological Superconductivity in Monolayer Transition Metal Dichalcogenides}
\author{Wen-Yu He}
\affiliation{Department of Physics, Hong Kong University of Science and Technology, Clear Water Bay, Hong Kong, China}
\author{Benjamin T. Zhou}
\affiliation{Department of Physics, Hong Kong University of Science and Technology, Clear Water Bay, Hong Kong, China}
\author{James J. He}
\affiliation{RIKEN Center for Emergent Matter Science (CEMS), Wako, Saitama 351-0198, Japan}
\author{Noah F. Q. Yuan}
\affiliation{Department of Physics, Massachusetts Institute of Technology, Cambridge, Massachusetts 02139, USA}
\author{Ting Zhang}
\affiliation{Department of Physics, Hong Kong University of Science and Technology, Clear Water Bay, Hong Kong, China}
\author{K. T. Law} \thanks{Corresponding author.\\phlaw@ust.hk}
\affiliation{Department of Physics, Hong Kong University of Science and Technology, Clear Water Bay, Hong Kong, China}

\begin{abstract}
Recently, Ising superconductors which possess in-plane upper critical fields much larger than the Pauli limit field are under intense experimental study. Many monolayer or few layer transition metal dichalcogenides are shown to be Ising superconductors. In this work, we show that in a wide range of experimentally accessible regimes where the in-plane magnetic field is higher than the Pauli limit field but lower than $H_{c2}$, a 2H-structure monolayer NbSe$_2$ or simiarly TaS$_2$ becomes a nodal topological superconductor. The bulk nodal points appear on the $\Gamma- M$ lines of the Brillouin zone where the Ising SOC vanishes. The nodal points are connected by Majorana flat bands, similar to the Weyl points being connected by surface Fermi arcs in Weyl semimetals. The Majorana flat bands are associated with a large number of zero energy Majorana fermion edge modes which induce spin-triplet Cooper pairs. This work demonstrates an experimentally feasible way to realise Majorana fermions in nodal topological superconductor, without any fining tuning of experimental parameters.

\end{abstract}

\maketitle

\section{Introduction}

In recent experiments, a new type of superconductors called Ising superconductors were discovered. It was shown that monolayer superconducting transition metal dichalcogenides (TMD) such as MoS$_2$~\cite{Ye2, Iwasa2} and NbSe$_2$~\cite{Mak2}  possess extremely high in-plane upper critical field $H_{c2}$, more than six times higher than the Pauli limit. It was explained that the enhancement of $H_{c2}$ is due to the presence of a special type of SOC called Ising SOC~\cite{Ye2, Mak2, King}. Ising SOC is caused by the lattice structure of monolayer TMD which breaks an in-plane mirror symmetry and it pins electrons spins to the out-of-plane directions. This is in sharp contrast to the Rashba SOC which pins electron spins to in-plane directions~\cite{Rashba0, Sigrist}. While it is intriguing that Ising SOC can enhance the upper critical fields of superconductors, it is not understood whether Ising SOC can induce novel superconducting phases.

\begin{figure}
\centering
\includegraphics[width=3.5in]{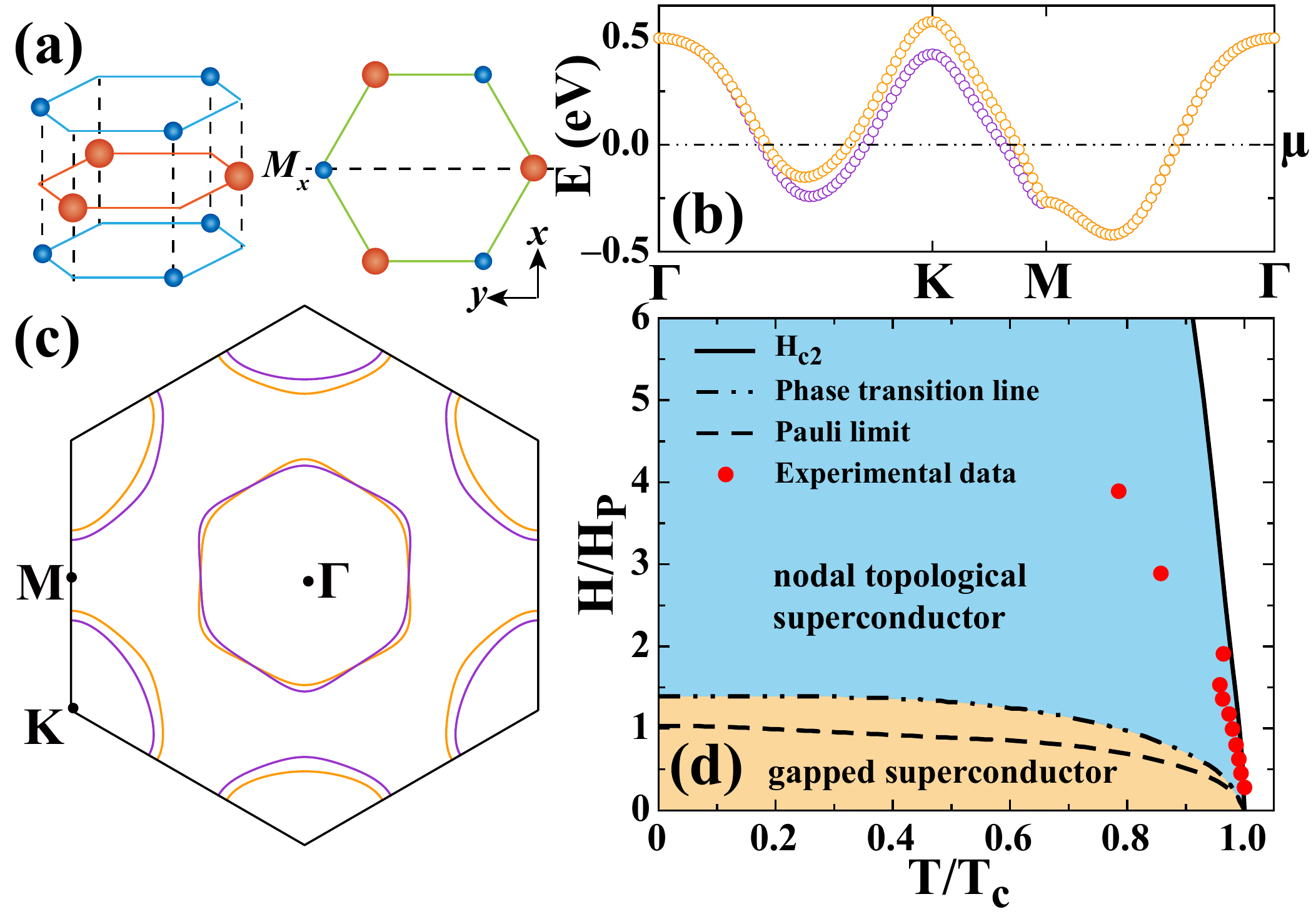}
\caption{ (a) Lattice structure of monolayer NbSe$_2$ (left) and its top view (right). $M_x$ denotes the in-plane mirror symmetry. (b) The band structure for the normal state of monolayer NbSe$_2$ from DFT calculations. (c) The energy contour at the Fermi level from tight-binding model. (d) The H-T phase diagram of the monolayer NbSe$_2$. The red dots represent the experimental data from Ref.~\cite{Mak2}. The calculated $H_{c2}$ is denoted by the solid line and it is much higher than the Pauli limit which is denoted by the dashed line. When the applied in-plane magnetic field is higher than the Pauli limit but lower than $H_{c2}$, the system can go through a phase transition from a fully gapped superconducting phase to a nodal topological phase.}
\label{FIG01}
\end{figure}

In this work, we show that, due to the strong Ising SOC and the fact that the in-plane $H_{c2}$ of superconducting TMDs is strongly enhanced, a superconducting monolayer TMD can be easily driven to the nodal topological phase by an in-plane magnetic field. As shown in Fig.1d, using monolayer NbSe$_2$ as an example, that the nodal topological phase occupies a large part of the phase diagram in Fig.1d and the predicted nodal topological regime has been achieved in recent experiments~\cite{Mak2,Pasupathy, WangJian}.  As shown in Fig.2a, nodal points on $\Gamma$-M lines are created by an in-plane magnetic field and these nodal points are connected by Majorana flat bands as shown in Fig.2b. We further show that the Majorana flat bands can be detected by tunnelling experiments as illustrated in Fig.3. Recent experiments on semiconductor nanowire/superconductor heterostructures~\cite{ZhangHao} and quantum anomalous Hall/superconductor heterostructures~\cite{WangKang} have provided evidence of Majorana fermions in 1D and 2D fully gapped topological superconductors~\cite{Oppen, Sarma1, Alicea2, Law0}. In these systems, fine tuning of experimental parameters is needed to achieve the topological regime as the topological regimes are extremely narrow. This work demonstrates an experimentally feasible way to realise Majorana fermions in nodal topological superconductor, without engineering heterostructures and without any fining tuning of experimental parameters.

Moreover, the appearance of the nodal topological phase predicted does not depend on the detailed band structure of the system. It only depends on the fact that the Fermi surfaces near the $\Gamma$ point of the superconductors are split by Ising SOC. Our study applies to a whole class of TMDs such as monolayer superconducting TaS$_2$ ~\cite{Coronado, Xiaoming} and NbSe$_2$~\cite{Mak2} which are under intense experimental studies recently due to the coexistence of superconductivity with charge density way and strong Ising SOC.

In the following sections, we first present a realistic band structure of a monolayer NbSe2. Secondly, using the tight-binding model, we explain the origin of the enhanced in-plane $H_{c2}$ in NbSe$_2$. Thirdly, we discuss how the nodal topological phase can be induced by an in-plane magnetic field resulting in nodal points which are connected by Majorana flat bands. Finally, we study how to detect this topological nodal phase by tunnelling experiments.

\section{Results}

\subsection{\bf {Band structure in normal state}}
A monolayer NbSe$_2$ is formed by a layer of Nb atoms with triangular lattice sandwiched by two layers of Se atoms with triangular lattices. The material has a hexagonal lattice structure when viewed from the out-of-plane direction but with broken A-B sublattice symmetry as shown in Fig.\ref{FIG01}a. Therefore, an in-plane mirror symmetry along y direction is broken and this gives rise to the Ising SOC which pins electron spins to the out-of-plane directions and split the energy bands~\cite{Noah0}. The band structure of a monolayer NbSe$_2$ is obtained through first-principle calculations taking into account SOC using \emph{ABINIT} package~\cite{ABINIT}. The results are presented in Fig.\ref{FIG01}b. As expected, the shape of the band structure is almost identical to the band structures of monolayer MoS$_2$, MoSe$_2$, WSe$_2$ and WTe$_2$ found in previous works~\cite{Yao}. However, unlike Mo and W based materials which are insulating intrinsically, a Nb atom has one less $d$-electron in the outer most shell than Mo and W atoms and the chemical potential of NbSe$_2$ lies in the valence band. The band splitting at the K points due to Ising SOC is about 150meV. Importantly, the band splitting is significant at the Fermi energy even the Fermi surface is far away from the K points.  As we show in the next section, this Ising SOC at the Fermi energy plays a crucial role in protecting the superconductivity from the paramagnetic effects of in-plane magnetic fields. 

In order to study the superconducting properties of the material, we construct a six-band tight-binding model which consists of the  $d_{z^2}$ , $d_{xy}$ and $d_{x^2-y^2}$ orbitals~\cite{Yao, Eriksson} of the Nb atoms. Including the third-nearest-neighbor hoppings, this band structure of the DFT calculations can be well explained. The details of the tight-binding model $H_{N}({\bm k})$ and the fitting parameters can be found in the Appendix and the Supplementary Material~\cite{supplementary}. The Fermi surfaces from the tight-binding model are shown in Fig.\ref{FIG01}c. At the Fermi level, there are Fermi pockets around $\Gamma$, K and -K points. In general, the bands are split by Ising SOC except for the states lying along the $\Gamma-M$ lines. It is important to note that the Fermi surfaces of NbSe$_2$ as depicted in Fig.1c have been observed through angle-resolved photoemission spectroscopy ~\cite {Crommie, King}. 

\subsection{Ising superconductivity in monolayer NbSe$_2$}
It is well-known that bulk NbSe$_2$ is superconducting with $T_c$ of 7K~\cite{Beernsten}. However, monolayer NbSe$_2$ was shown to be superconducting only very recently with $T_c$ on-set at about 5K and zero resistance appears at about 3K~\cite{Mak1, Mak2, Crommie, Pasupathy}. In bulk NbSe$_2$, magnetic field can create vortices which eventually destroy superconductivity with $H_{c2}$ of about 14.5T ~\cite{Toyota}. In the monolayer case, the orbital effects of in-plane magnetic fields are completely suppressed and the magnetic field can suppress superconductivity only through Zeeman coupling with electron spins. 

Similar to the gated superconducting MoS$_2$, NbSe$_2$ was shown to have strongly enhanced $H_{c2}$ at 35T even though the $T_c$ is much lower than the bulk $T_c$. While the experimental data in MoS$_2$ can be well explained by solving the self-consistence gap equations by taking into account the Ising SOC and Rashba SOC induced by gating, a microscopic theory for explaining the $H_{c2}$ data of monolayer NbSe$_2$ is still lacking due to the more complicated band structures in the valence bands. In this work, with the tight-binding model introduced in the Appendix, we can take into account all the three Fermi pockets shown in Fig.\ref{FIG01}c to explain the experimental data. An intra-orbital attractive interaction $U$ ~\cite{supplementary} between the electrons is introduced. $U$ is determined by the self-consistent gap equation $\Delta=U/V\sum_{\bm{k},o}\left \langle\ c_{\bm{k},\downarrow,o}c_{-\bm{k},\uparrow,o}\right \rangle$ at zero magnetic field. Here, $\Delta = 1.76k_{B}T_c$ is the pairing gap at zero temperature, $V$ is the area of the sample and $c_{\bm{k},\uparrow/\downarrow,o}$ is the electron annihilation operator for orbital $o$.

After finding $U$ at zero magnetic field, we can determine $\Delta$ as a function of magnetic field by minimizing the free energy density
\begin{eqnarray}
\Omega=\left |\Delta\right |^2/U-\frac{1}{V}\sum_{\bm{k},n}\frac{1}{2\beta}\ln\left(1+e^{-\beta E_{\bm{k},n}}\right)
\end{eqnarray}
with $E_{\bm{k},n}$ the eigenvalue adopted from the Bogoliubov-de Gennes Hamiltonian $H_{\text{BDG}}$ as shown in detail in the Supplementary Material~\cite{supplementary}. The phase diagram for this Ising superconductor is determined and shown in Fig.\ref{FIG01}(d), with $H_p\approx{1.84}T_c$ the Pauli limit~\cite{Clogston}. As shown in Fig.\ref{FIG01}(d), the calculated $H_{c2}$ (solid line) is strongly enhanced. The theoretical values of $H_{c2}$, using the parameters from the tight-binding model without any tuning parameters, are higher than the experimental values (the red dots). Such discrepancy can be due to the fact that the NbSe$_2$ sample has ripples ~\cite{Fai}, and the ripples introduce Rashba type SOC ~\cite{Hernando} which competes with Ising SOC. This Rashba type SOC can lower the in-plane $H_{c2}$ as in the case of gated superconducting MoS$_2$~\cite{Ye2}. We ignore the effects of ripples in this work. To highlight the importance of Ising SOC, the $H_{c2}$ in the absence of Ising SOC is denoted by the dashed line in Fig.\ref{FIG01}(d). It is evident that the Ising SOC enhances $H_{c2}$ so that it becomes much higher than the Pauli limit.
\begin{figure}
\centering
\includegraphics[width=3.5in]{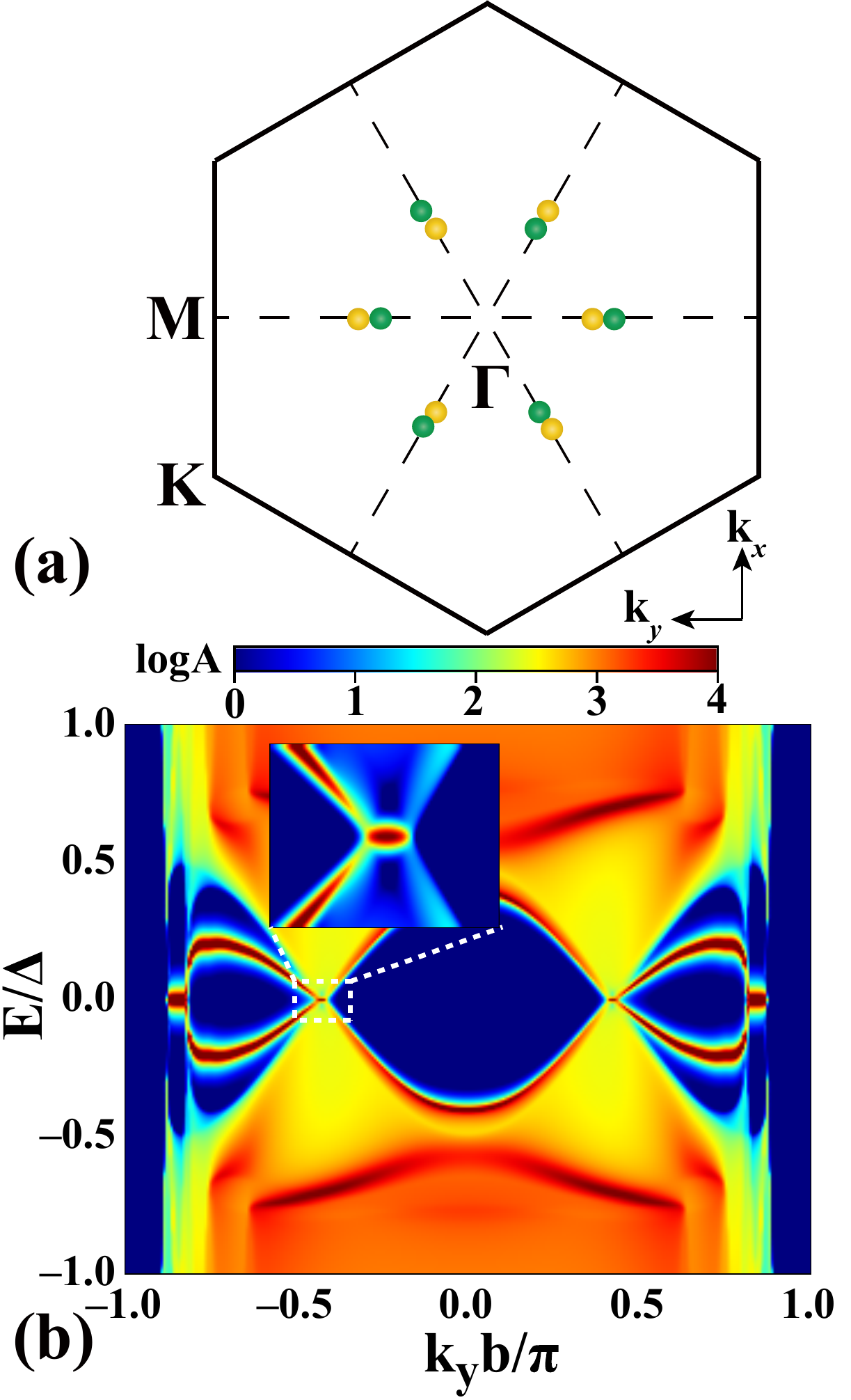}
\caption{The nodal topological superconducting monolayer NbSe$_2$ driven by in-plane magnetic field with $H=3H_{P}$. (a) The position for the six pairs of point nodes. They project into four pairs of nodal points on the armchair edges as shown in (b).  (b) The logarithmic plot of the spectral function $A(E,\bm{R})$ on the armchair edge a semi-infinite strip of NbSe$_2$ with open boundary condition in the $x$-direction and periodic boundary condition in the $y$-direction. The Majorana flat bands connecting the point nodes at the sample edge. The insert shows a pair of nodal points connected by a Majorana flat band. Here $b=\sqrt{3}a$ and $a$ the lattice constant}
\label{FIG02}
\end{figure}

\subsection{Nodal topological phase}
From Fig.\ref{FIG01}d, one can see that superconductivity survives in the regime where the applied magnetic field is higher than the Pauli limit field. The question is: what are the properties of the superconducting phase under strong magnetic fields? To answer this question, the spectral function $A(E, \bm{R})=-\frac{1}{\pi}\text{Tr}\left[\text{Im}G\left(E, \bm{R}\right)\right]$ of the a semi-infinite NbSe$_2$ stripe is shown in Fig.\ref{FIG02}b. Here, $E$ is energy and $\bm{R}$ is a point on the armchair edge and $G$ is the Green's function of the BdG Hamiltonian defined in Supplementary Material~\cite{supplementary}. The armchair edges are parallel to the $y$-direction and subject to periodic boundary conditions. The in-plane magnetic field applied is chosen to be $H = 3 H_p$. It can be seen that there are four sections of Majorana flat bands in Fig.\ref{FIG02}a. The Majorana flat bands connect the four pairs of bulk nodal points. These four pairs of nodal points are projections of the six pairs of bulk nodal points in the bulk energy spectrum as shown in Fig.\ref{FIG02}a.

To understand this nodal topological phase, we use an effective Hamiltonian approach and understand the system from a symmetry point of view. Before writing down the effective Hamiltonian near the Fermi energy, we note that the Ising SOC near the K points is very strong as shown in Fig.\ref{FIG01}b. Therefore, the superconducting states near K points are hardly affected by the in-plane magnetic field at $H = 3 H_p \ll H_{c2}$. Therefore, to understand the gap closing effect of the in-plane magnetic field, we only need to focus on the states in the $\Gamma$ pocket where Ising SOC is weaker.

Around the $\Gamma$ pocket, the $d_{z^2}$ orbital dominates~\cite{Eriksson, Yao}. In the basis of $[c_{k, \uparrow}, c_{k, \downarrow}]$ and in the absence of external magnetic fields, the Hamiltonian has to satisfy the point group symmetries $M_z=-i\sigma_z$, $M_x=-i\sigma_x$, $C_3=e^{-i\frac{\pi}{3}\sigma_z}$ and time-reversal symmetry $T=i\sigma_yK$. These symmetries restrict the effective Hamiltonian, up to third order in $k$, to have the form:
\begin{equation}
\mathcal{H}_{0}= [\frac{k_x^2+k_y^2}{2m}-\mu]\sigma_0+\lambda_{\text{SOC}}\left(k^3_{+}+k^3_{-}\right)\sigma_z .
\end{equation}

Here, $\mu$ is the chemical potential and $k_{\pm} = k_x \pm i k_y$. It is important to note that $\lambda_{\text{SOC}}$ term vanishes along the $\Gamma-M$ lines as dictated by symmetries, consistent with the calculations in Fig.\ref{FIG01}c.

Including the in-plane magnetic field, spin-singlet pairing, and in the basis of $[ c_{k, \uparrow}, c_{k, \downarrow}, c_{-k, \uparrow}^{\dagger}, c_{-k, \downarrow}^{\dagger}$], the Hamiltonian in the superconducting phase can be written as:
\begin{eqnarray}
\mathcal{H}_{s}&=&\left[\frac{k_x^2+k_y^2}{2m}-\mu\right]\tau_z+\lambda_{\text{SOC}}\left(k^3_{+}+k^3_{-}\right)\sigma_z\\\nonumber&&+\frac{1}{2}\mu_BgH_x\tau_z\sigma_x+\frac{1}{2}\mu_BgH_y\sigma_y + \Delta\tau_y\sigma_y.
\end{eqnarray}
Here, $\tau_i$ denotes Pauli matrices in the particle-hole basis. $H_x$ and $H_y$ are magnetic fields in the x- and y-directions respectively, $\mu_B$ is the Bohr magneton and $g$ is the electron's gyromagnetic ratio. The pairing potential is denoted by $\Delta$. Even though time-reversal is broken by the external magnetic field, $\mathcal{H}_{s}$ respects a time-reversal like symmetry $U_T\mathcal{H}_{s}\left(k_x, k_y\right)U_T^{-1}=\mathcal{H}_{s}\left(-k_x, k_y\right)$ with $U_T=M_zT\tau_z$ and $k_y$ unchanged under the symmetry operation. Moreover, $\mathcal{H}_{s}$ respects a 1D particle-hole like symmetry $U_P\mathcal{H}_{s}\left(k_x, k_y\right)U_P^{-1}=-\mathcal{H}_{s}\left(-k_x, k_y\right)$ with $U_P=\tau_xK$. Therefore, for fixed $k_y$, $\mathcal{H}_{s}$ respects the chiral symmetry $C\mathcal{H}_{k_y}(k_x) C^{-1} = -\mathcal{H}_{k_y}(k_x)$ where $C=\sigma_x\tau_y$. As a result, for any fixed $k_y$, $\mathcal{H}_{s}$ is in the BDI class and the Hamiltonian can be topologically non-trivial ~\cite{Schnyder, Sau, Law, Noah}. For the range of $k_y$ where $\mathcal{H}_{s}$ is non-trivial, there are Majorana zero energy modes on the edge of the system as shown in Fig.\ref{FIG02}b. By tuning $k_y$ as a parameter, the system can undergo a topological phase transition from a trivial regime to a non-trivial regime by closing the bulk gap. These topological phase transition $k_y$ points are the nodal points in Fig.\ref{FIG02}b. The nodal points and the Majorana flat bands shown in Fig.2 can be reproduced by the simple Hamiltonian $H_{s}$ in Eq.3.

\subsection{Detection}
In this section, we discuss the experimental detection of the nodal topological superconducting phase in NbSe$_2$. As discussed above, superconducting NbSe$_2$ can be driven from a fully gapped superconducting phase to a nodal superconducting phase by an in-plane magnetic field. As in the case of $d_{x^2-y^2}$-wave superconductors, we expect the density of states of the system as a function of energy to be $V$-shape near zero energy when the system is nodal. From the low-energy bulk spectrum of the effective Hamiltonian $\mathcal{H}_{s}$, the density of states near zero energy can be found to be $N\frac{2\pi \left |E\right |}{v_x v_y}$, where $E$ is the energy and $v_x,v_y$ denote the Fermi velocity in the $x,y$-directions near the nodal points and $N$ is the number of nodal points. Thus, the bulk density of states near zero energy is indeed linearly proportional to $\left |E\right|$.

\begin{figure}
\includegraphics[width=3.6in]{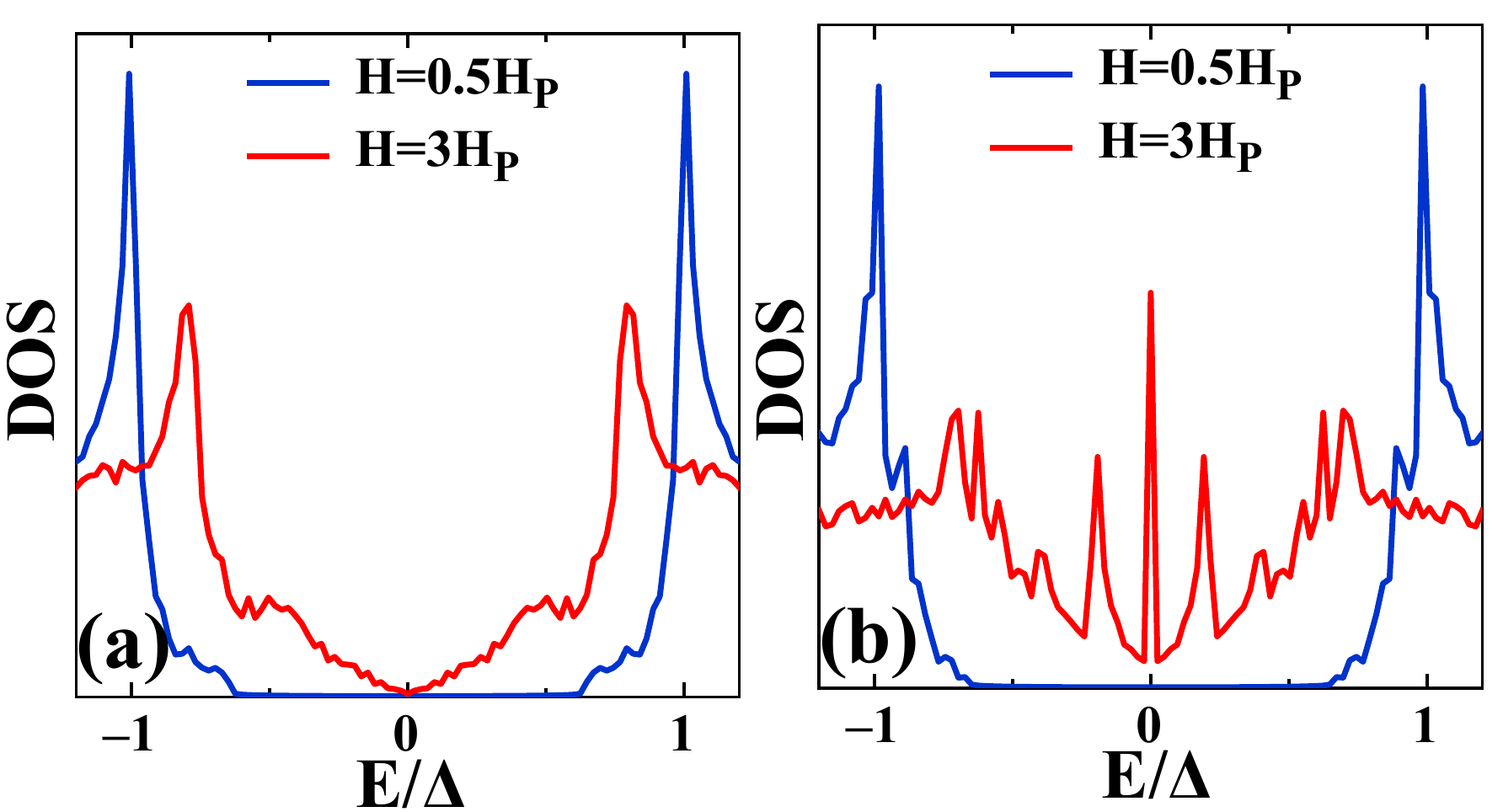}
\caption{ (a) The local density of states (DOS) in the bulk in the fully gapped regime (blue curve) and the nodal regime (red curve) respectively. The in-plane magnetic field changes the DOS from U-shape ($H= 0.5H_{P}$ to V-shape $H= 3H_{P}$. (b) The DOS on the armchair edge in the nodal topological phase with $H= 3H_{P}$. The zero energy DOS is strongly enhanced due to the zero energy Majorana modes associated with the Majorana flat bands (red curve).}
\label{FIG03}
\end{figure}

Using the real space version of the six-band tight-binding model $H_{\text{BDG}}$ defined in the Appendix and the Supplementary Material~\cite{supplementary}, including pairing and the in-plane magnetic field, we calculate the local density of states in the bulk of the sample. 
\begin{equation}
\rho\left(E\right)=-\frac{1}{\pi}\text{Tr}\left[\text{Im} G^{\text{R}}\left(E\right)\right],
\end{equation}
here $G^{\text{R}}\left(E\right)=\left(E+i\eta-H_{\text{BdG}}\right)^{-1}$ is the retarded Green's function. The local density of states for a point in the bulk is shown in Fig.\ref{FIG03}a. When the applied in-plane field is lower than the Pauli limit, the density of states as a function of energy is $U$-shape (blue curve) near zero energy indicating a fully gapped superconducting state. On the other hand, the density of states is changed to $V$-shape (red curve) when the system is driven to the nodal phase as expected. Even though the $V$-shape density of states in the bulk is only the signature of a nodal phase, not necessarily a nodal topological phase, the observation of the fully gapped to nodal phase transition can be important for identifying the topological phase transition point.

Moreover, in the nodal topological phase, there are a large number of zero energy Majorana modes residing on the edges of the sample. These zero energy modes strongly enhance the local density of states on the edge of the superconductor. The local density of states at the edge of the sample, before and after the topological phase transition, are shown in Fig.\ref{FIG03}b. It is evident that, in the nodal topological phase with in-plane field stronger than the Pauli limit field, the zero energy density of states are strongly enhanced (red curve). Both the bulk density of states and the edge density of states can be detected by tunnelling spectroscopy experiments.

\begin{figure}
\centering
\includegraphics[width=3.5in]{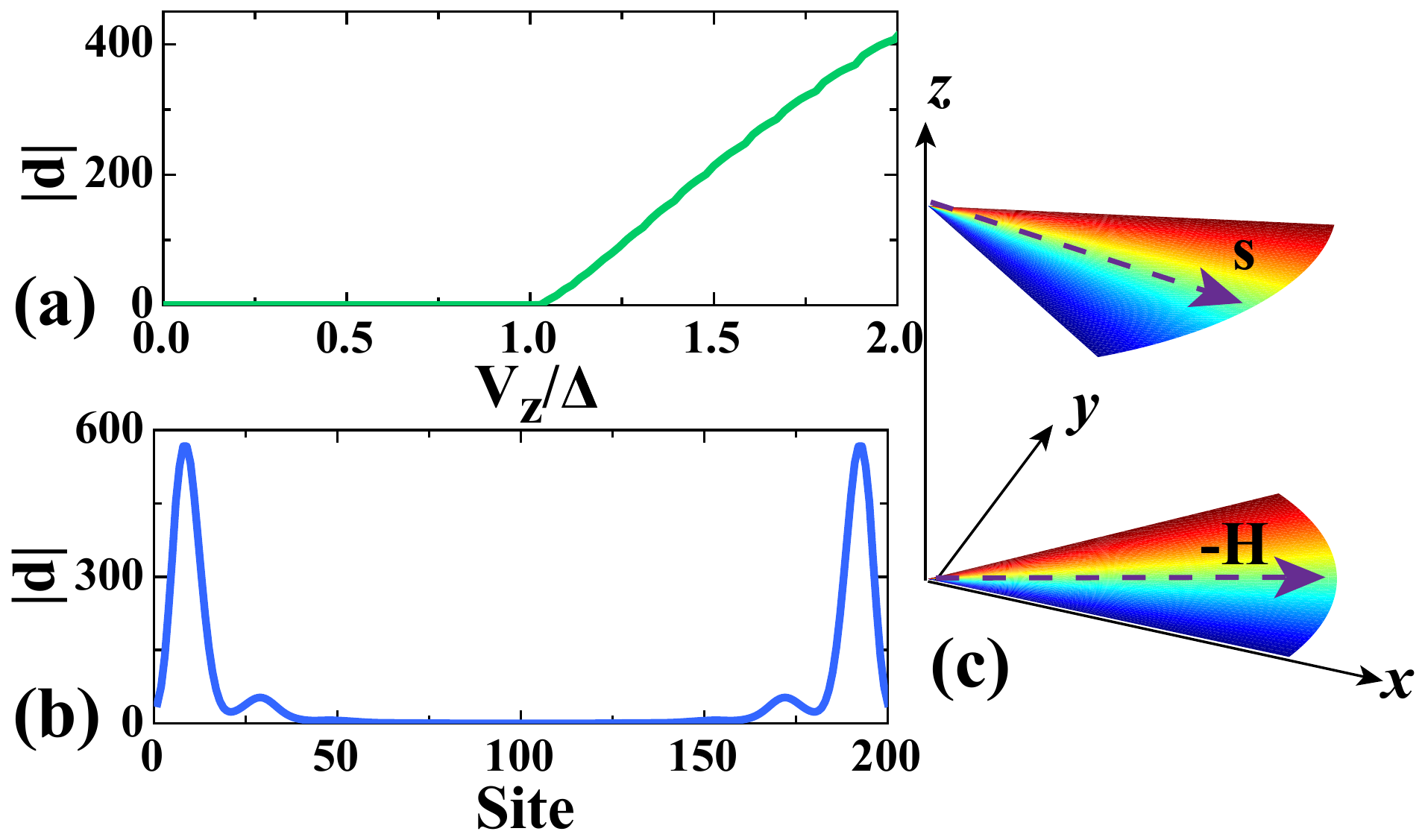}
\caption{(a) The triplet-pairing correlation amplitude $|\bm{d}|$ on the armchair edge of the sample, as a function of $V_z/\Delta$ where $V_z=\frac{1}{2}\mu_BgH$. It becomes nonzero after the Majorana flat bands emerge when $V_z > \Delta$. (b) The spatial dependence of $|\bm{d}|$ in the nodal topological phase with $H=3H_P$. Here the site are along the $x$-direction. (c) The spin polarization direction $\bm{s}$ of the Cooper pair at the armchair edge. $\bm{s}$ depends on the direction of the applied magnetic field $\bm{H}$. The vector $\bm{s}$ with certain color is determined by the vector $-\bm{H}$ of the same color. }
\label{FIG04}
\end{figure}

Interestingly, due to the zero energy Majorana modes, the pairing correlations induced on the edge of the superconductor are indeed spin-triplet. To show this, we calculated the pairing correlation for a site on the armchair edge of the sample. The pairing correlation can be parametrized as ~\cite{Rashba, Sigrist, Zhou}:
\begin{eqnarray}
F_{\sigma\sigma'}\left(E\right)=-i\sum_{o}\int_{0}^{\infty}e^{i\left(E+i0^{+}\right)t}\left \langle\left\{c_{\sigma,o}\left(t\right), c_{\sigma',o}\left(0\right)\right\} \right \rangle dt
\end{eqnarray}
where $\sigma$ and $o$ are the spin and orbital indices respectively. In the matrix form, the pairing correlation is written as:
\begin{eqnarray}
F\left(E\right)=\left(\psi+\bm{d}\cdot\bm{\sigma}\right)i\sigma_y.
\end{eqnarray}
Here, the so-called $\bm{d}$-vector characterizes the triplet-pairing correlation and $\psi$ characterizes the spin-singlet pairing correlation. In Fig.\ref{FIG04}a, it is shown that the triplet pairing correlation at zero energy on the armchair edge of the sample is non-zero only when the Zeeman energy $V_z= \frac{1}{2}\mu_{B}g |H|$ is larger than $\Delta$. Fig.\ref{FIG04}b shows the spatial variation of $| \bm{d}|$ across the sample for strip of NbSe$_2$ which has a width of 200-sites in the $x$-direction (perpendicular to the armchair edge). It is evident that $|\bm{d}|$ is significant only near the two edges of the sample. On the other hand, the singlet pairing correlation amplitude $\psi$ at zero energy is negligible. Furthermore, we calculated $\bm{s}=i\left(\bm{d}\times\bm{d}^{\ast}\right)/|\bm{d}|^2$ which gives the spin-polarization direction of the Cooper pairs~\cite{Leggett} as shown in Fig.\ref{FIG04}c. This indicates that the Cooper pairs injected into the superconductor by the normal lead are spin-polarized with spin pointing to $\bm{s}$-direction~\cite{Noah, James}. As a result, even though the lead attached to the superconductor is non-magnetic, the current from the lead to the superconductor is spin-polarized. In other words, the Majorana modes can act as spin filters and only allow electrons with spin pointing to $\bm{s}$-direction to tunnel into the superconductor. We believe this property of monolayer NbSe$_2$ may lead to applications in superconducting spintronics~\cite{Linder, Eschrig}.

\section{Discussion}

We would like to discuss some important points about the nodal topological phase mentioned above. First, the large number of Majorana zero energy modes associated with the Majorana flat band in Fig.2b are protected against disorder. This is in sharp contrast to the $d_{x^2-y^2}$-wave superconductor in which the zero energy fermionic modes can be lifted to finite energy by disorder. One can show that the zero energy Majorana modes is not protected by the bulk gap (since the system is nodal) but by the chiral symmetry $C=\sigma_x\tau_y$ of $H_{s}$ in Eq.3 ~\cite{Yokoyama}. This chiral symmetry is not broken by on-site disorders. This is similar to the toy model studied in Ref.~\cite{Law}, where an in-plane magnetic field creates Majorana flat bands in a time-reversal invariant $p \pm ip$ topological superconductor.

Second, we emphasis that the $\Gamma$ pocket is essential for creating the nodal topological phase. In the case of gated MoS$_2$ or WS$_2$, superconductivity appear in the conduction bands near the $K$ pockets only. Therefore, the nodal phase discussed above cannot be found in those superconductors. On the other hand, intrinsic superconductors NbSe$_2$, NbS$_2$, TaSe$_2$ and TaS$_2$ are candidate materials for realizing the nodal topological superconducting phase found in this work.

Third, the armchair edge is not essential for the observation of the Majorana flat bands as long as the edge is not parallel to the zig-zag edge ($k_x$ direction in Fig.2a). On the zig-zag edge, the projection of the bulk nodal points can cancel each other such that there are no topologically non-trivial regimes for all $k_x$ parallel to the zig-zag edge. For all other edges, one can find finite sections of Majorana flat bands along the edges when the system is driven to the nodal topological phase.

\section*{Acknowledgement}
The authors thank Patrick A. Lee, and Masatoshi Sato for insightful discussions. W. -Y. H acknowledges the support of Hong Kong PhD Fellowship. K. T. L thanks the support of HKRGC and Croucher Foundations through HKUST3/CRF/13G,
602813, 605512, 16303014 and Croucher Innovation Grants.

\appendix*
\section{\bf {Tight-binding Hamiltonians}}\label{App}

In the basis of $[c_{\bm{k},d_{z^2},\uparrow}$, $c_{\bm{k}, d_{xy}, \uparrow}$, $c_{\bm{k}, d_{x^2-y^2},\uparrow}$, $c_{\bm{k},d_{z^2},\downarrow}$, $c_{\bm{k}, d_{xy}, \downarrow}$, $c_{\bm{k}, d_{x^2-y^2},\downarrow}]$, the six-band model used has the form~\cite{Yao}
\begin{align}\nonumber
H_{\text{N}}\left(\bm{k}\right)& =\sigma_0\otimes H_{\text{TNN}}\left(\bm{k}\right)+\sigma_z\otimes\frac{1}{2}\lambda L_z\\
&\quad+\frac{1}{2}\mu_Bg\bm{H}\cdot\bm{\sigma}\otimes I_3
\end{align}
with
\begin{eqnarray}
H_{\text{TNN}}\left(\bm{k}\right)=\begin{pmatrix}
V_0 & V_1 & V_2 \\ 
V_1^{\ast} & V_{11} & V_{12} \\ 
V_2^{\ast} & V_{12}^{\ast} & V_{22}
\end{pmatrix}, L_z=\begin{pmatrix}
0 & 0 & 0 \\ 
0 & 0 & -i \\ 
0 & i & 0
\end{pmatrix}
\end{eqnarray}
where $I_n$ means the $n\times n$ identity matrix. The details about these matrix elements can be found in supplementary material~\cite{supplementary}. Then the Bogliubov-de Gennes Hamiltonian in Nambu spinor representation can be written as
\begin{eqnarray}
H_{\text{BdG}}\left(\bm{k}\right)=\begin{pmatrix}
H_{\text{N}}\left(\bm{k}\right)-\mu I_6 & -i\Delta\sigma_y\otimes I_3\\ 
i\Delta\sigma_y\otimes I_3 & -H^{\ast}_{\text{N}}\left(\bm{-k}\right)+\mu I_6
\end{pmatrix}.
\end{eqnarray}
The real space version of $H_{\text{TNN}}$ and $H_{\text{BdG}}$ can be found in the Supplementary Material~\cite{supplementary}.

\onecolumngrid
\clearpage
\begin{center}
{\bf Supplementary Material: Magnetic Field Driven Nodal Topological Superconductivity in Monolayer Transition Metal Dichalcogenides}
\end{center}
	
\author{{ Wen-Yu He, Benjamin T. Zhou, James J. He, Noah F. Q. Yuan, Ting Zhang, and K. T. Law}\\
{\small \em Department of Physics, Hong Kong University of Science and Technology, Clear Water Bay, Hong Kong, China.\\}}
\maketitle
\setcounter{equation}{0}
\setcounter{figure}{0}
\setcounter{table}{0}
\setcounter{page}{1}
\makeatletter
\renewcommand{\theequation}{S\arabic{equation}}
\renewcommand{\thefigure}{S\arabic{figure}}
\renewcommand{\bibnumfmt}[1]{[S#1]}
\renewcommand{\citenumfont}[1]{S#1}
	
\section{Tight binding model}
The valence band of monolayer NbSe$_2$ is dominated by the $d_{z^2}$, $d_{xy}$ and $d_{x^2-y^2}$ orbitals from Nb atoms~\cite{Eriksson, Yao}. In the basis of $[c_{\bm{k},d_{z^2},\uparrow}, c_{\bm{k}, d_{xy}, \uparrow}, c_{\bm{k}, d_{x^2-y^2},\uparrow}, c_{\bm{k},d_{z^2},\downarrow}, c_{\bm{k}, d_{xy}, \downarrow}, c_{\bm{k}, d_{x^2-y^2},\downarrow}]$, the corresponding normal state tight binding model up to the third-nearest-neighbor hopping can be written as
\begin{eqnarray}
H_{\text{N}}\left(\bm{k}\right)=\sigma_0\otimes H_{\text{TNN}}\left(\bm{k}\right)+\sigma_z\otimes\frac{1}{2}\lambda L_z+\bm{H}\cdot\bm{\sigma}\otimes I_3
\end{eqnarray}
with
\begin{eqnarray}
H_{\text{TNN}}\left(\bm{k}\right)=\begin{pmatrix}
V_0 & V_1 & V_2 \\ 
V_1^{\ast} & V_{11} & V_{12} \\ 
V_2^{\ast} & V_{12}^{\ast} & V_{22}
\end{pmatrix}, L_z=\begin{pmatrix}
0 & 0 & 0 \\ 
0 & 0 & -i \\ 
0 & i & 0
\end{pmatrix}
\end{eqnarray}
where $I_3$ is the $3\times3$ identity matrix. Defining $\left(\alpha, \beta\right)=\left(\frac{1}{2}k_xa, \frac{\sqrt{3}}{2}k_ya\right)$, $V_0$, $V_1$, $V_2$, $V_{11}$, $V_{12}$ and $V_{22}$ can be expressed as
\begin{eqnarray}
V_0=\epsilon_1+2t_0\left(2\cos\alpha\cos\beta+\cos 2\alpha\right)+2r_0\left(2\cos3\alpha\cos\beta+\cos2\beta\right)+2u_0\left(2\cos2\alpha\cos2\beta+\cos4\alpha\right),
\end{eqnarray}
\begin{eqnarray}
\text{Re}\left[V_1\right]=-2\sqrt{3}t_2\sin\alpha\sin\beta+2\left(r_1+r_2\right)\sin3\alpha\sin\beta-2\sqrt{3}u_2\sin2\alpha\sin2\beta,
\end{eqnarray}
\begin{eqnarray}
\text{Im}\left[V_1\right]=2t_1\sin\alpha\left(2\cos\alpha+\cos\beta\right)+2\left(r_1-r_2\right)\sin3\alpha\cos\beta+2u_1\sin2\alpha\left(2\cos2\alpha+\cos2\beta\right),
\end{eqnarray}
\begin{eqnarray}
\text{Re}\left[V_2\right]=2t_2\left(\cos2\alpha-\cos\alpha\cos\beta\right)-\frac{2}{\sqrt{3}}\left(r_1+r_2\right)\left(\cos3\alpha\cos\beta-\cos2\beta\right)+2u_2\left(\cos4\alpha-\cos2\alpha\cos2\beta\right),
\end{eqnarray}
\begin{eqnarray}
\text{Im}\left[V_2\right]=2\sqrt{3}t_1\cos\alpha\sin\beta+\frac{2}{\sqrt{3}}\left(r_1-r_2\right)\sin\beta\left(\cos3\alpha+2\cos\beta\right)+2\sqrt{3}u_1\cos2\alpha\sin2\beta,
\end{eqnarray}
\begin{eqnarray}
V_{11}=&&\epsilon_2+\left(t_{11}+3t_{22}\right)\cos\alpha\cos\beta+2t_{11}\cos2\alpha+4r_{11}\cos3\alpha\cos\beta+2\left(r_{11}+\sqrt{3}r_{12}\right)\cos2\beta\\
&&+\left(u_{11}+3u_{22}\right)\cos2\alpha\cos2\beta+2u_{11}\cos4\alpha,
\end{eqnarray}
\begin{eqnarray}
\text{Re}\left[V_{12}\right]=\sqrt{3}\left(t_{22}-t_{11}\right)\sin\alpha\sin\beta+4r_{12}\sin3\alpha\sin\beta+\sqrt{3}\left(u_{22}-u_{11}\right)\sin2\alpha\sin2\beta,
\end{eqnarray}
\begin{eqnarray}
\text{Im}\left[V_{12}\right]=4t_{12}\sin\alpha\left(\cos\alpha-\cos\beta\right)+4u_{12}\sin2\alpha\left(\cos2\alpha-\cos2\beta\right),
\end{eqnarray}
and
\begin{eqnarray}\nonumber
V_{22}=&&\epsilon_2+\left(3t_{11}+t_{22}\right)\cos\alpha\cos\beta+2t_{22}\cos2\alpha+2r_{11}\left(2\cos3\alpha\cos\beta+\cos2\beta\right)\\
&&+\frac{2}{\sqrt{3}}r_{12}\left(4\cos3\alpha\cos\beta-\cos2\beta\right)+\left(3u_{11}+u_{22}\right)\cos2\alpha\cos2\beta+2u_{22}\cos4\alpha.
\end{eqnarray}

\begin{figure}
\includegraphics[width=3.6in]{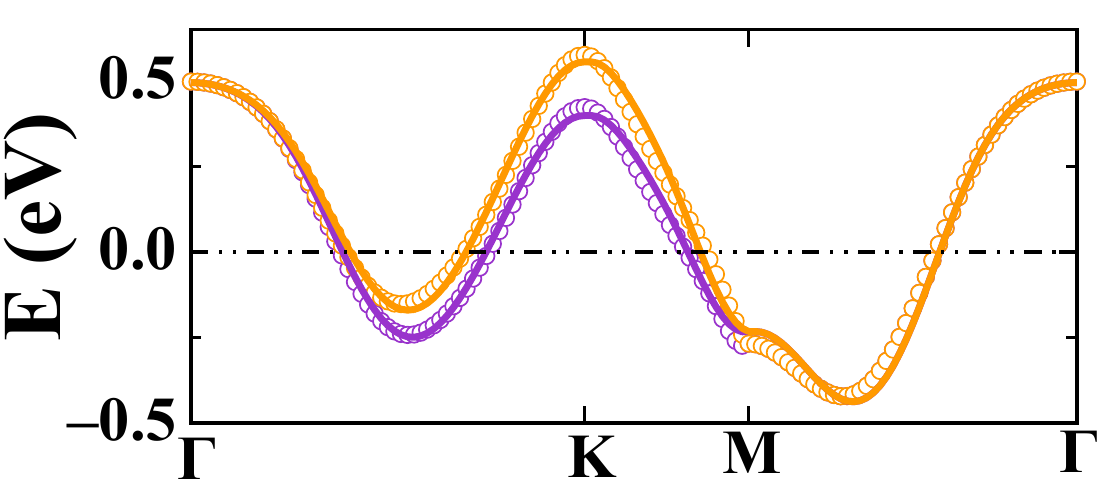}
\caption{The band structure for the normal state of monolayer NbSe$_2$. The circles represent the first principle calculation results, while the solid lines are the tight binding results.}
\label{FIGS1}
\end{figure}

\begin{table}[ht]
\caption{Fitting parameters for the Hamiltonian $H_{\text{TNN}}\left(\bm{k}\right)$. The energy parameters $\epsilon_1$ to $\lambda$ are in units of eV.} 
\centering 
\begin{tabular}{c c c c c c c c c c} 
\hline\hline 
$\epsilon_1$ & $\epsilon_2$ & $t_0$ & $t_1$ & $t_2$ & $t_{11}$ & $t_{12}$ & $t_{22}$ & $r_0$ & $r_1$\\
$r_2$ & $r_{11}$ & $r_{12}$ & $u_0$ & $u_1$ & $u_2$ & $u_{11}$ & $u_{12}$ & $u_{22}$ & $\lambda$\\ [0.5ex] 
\hline 
1.4466 & 1.8496 & -0.2308 & 0.3116 & 0.3459 & 0.2795 & 0.2787 & -0.0539 & 0.0037 & -0.0997 \\ 
0.0385 & 0.0320 & 0.0986 & 0.0685 & -0.0381 & 0.0535 & 0.0601 & -0.0179 & -0.0425 & 0.0784 \\ [1ex] 
\hline 
\end{tabular}
\label{table:nonlin} 
\end{table}
The specific parameters in this third-nearest-neighbor tight binding model can be fitted from the band structure by first principle and are summarized in Table I. It can be seen from Fig. \ref{FIGS1} that the third-nearest-neighbor tight binding model matches well with the first principle calculation results. Then the Bogliubov-de Gennes Hamiltonian in Nambu spinor representation can be written as
\begin{eqnarray}
H_{\text{BdG}}\left(\bm{k}\right)=\begin{pmatrix}
H_{\text{N}}\left(\bm{k}\right)-\mu \sigma_0 \otimes I_3 & -i\Delta\sigma_y\otimes I_3\\ 
i\Delta\sigma_y\otimes I_3 & -H^{\ast}_{\text{N}}\left(\bm{-k}\right)+\mu \sigma_0 \otimes I_3
\end{pmatrix}.
\end{eqnarray}

\section{Model for Armchair Nanoribbon}
In this section, we apply the third-nearest-neighbor tight binding model to study the superconducting monolayer NbSe$_2$ nano-ribbon. Taking the armchair direction along $y$, we have $N$ Nb atoms in one unit cell as shown in Fig. S2. With periodic boundary conditions in the $y$ direction and open boundary conditions in the $x$ direction, the $6N\times6N$ Hamiltonian matrix for the normal state armchair nano-ribbon can be obtained as follows
\begin{eqnarray}
H_{\text{ribbon}}\left(k_y\right)=I_2\otimes H_{0}+\sigma_z\otimes I_N\otimes\frac{1}{2}\lambda L_z+\bm{H}\cdot\bm{\sigma}\otimes I_{3N}.
\end{eqnarray}
where $I_n$ ($n=2, N, 3N$) means the $n\times n$ identity matrix and $H_{0}$ reads
\begin{eqnarray}
H_{0}=\begin{pmatrix}
h_1 & h_2 & h_3 & h_4 & h_5 & 0 & \cdots  & 0\\ 
h_2^{\dagger} & h_1  & h_2 & h_3 & h_4 & h_5 &  & \vdots\\ 
h_3^{\dagger} & h_2^{\dagger} & h_1 & h_2 & \ddots & \ddots & \ddots & 0\\ 
h_4^{\dagger} & h_3^{\dagger} & h_2^{\dagger} & \ddots & \ddots & \ddots & h_4 & h_5\\ 
h_5^{\dagger} & h_4^{\dagger} & \ddots & \ddots & \ddots & \ddots & h_3 & h_4\\ 
0 & h_5^{\dagger} & \ddots & \ddots & \ddots & h_1 & h_2 & h_3\\ 
\vdots &  & \ddots & h_4^{\dagger} & h_3^{\dagger} & h_2^{\dagger} & h_1 & h_2\\ 
0 & \cdots & 0 & h_5^{\dagger} & h_4^{\dagger} & h_3^{\dagger} & h_2^{\dagger} & h_1
\end{pmatrix}_{N\times N}.
\end{eqnarray}
Here $h_n$ ($n=1, 2, 3, 4, 5$), the hopping matrix between different sites reads
 \begin{eqnarray}
h_1=\left(\begin{smallmatrix}
\epsilon_1+2r_0\cos2\beta & 0 & \frac{2\left(r_1+r_2\right)}{\sqrt{3}}\cos2\beta+i\frac{2\left(r_1-r_2\right)}{\sqrt{3}}\sin2\beta \\ 
0 & \epsilon_2+2\left(r_{11}+\sqrt{3}r_{12}\right)\cos2\beta & 0\\ 
\frac{2\left(r_1+r_2\right)}{\sqrt{3}}\cos2\beta-i\frac{2\left(r_1-r_2\right)}{\sqrt{3}}\sin2\beta & 0 & \epsilon_2+\left(2r_{11}-\frac{2}{\sqrt{3}}r_{12}\right)\cos2\beta
\end{smallmatrix}\right),
\end{eqnarray}
\begin{eqnarray}
h_2=\left(\begin{smallmatrix}
2t_0\cos\beta & -i\sqrt{3}t_2\sin\beta-t_1\cos\beta &-t_2\cos\beta+i\sqrt{3}t_1\sin\beta \\ 
-i\sqrt{3}t_2\sin\beta+t_1\cos\beta & \frac{1}{2}\left(t_{11}+3t_{22}\right)\cos\beta &i\frac{\sqrt{3}}{2}\left(t_{22}-t_{11}\right)\sin\beta+2t_{12}\cos\beta \\ 
-t_2\cos\beta-i\sqrt{3}t_1\sin\beta & i\frac{\sqrt{3}}{2}\left(t_{22}-t_{11}\right)\sin\beta-2t_{12}\cos\beta &\frac{1}{2}\left(3t_{11}+t_{22}\right)\cos\beta
\end{smallmatrix}\right),
\end{eqnarray}
\begin{eqnarray}
h_3=\left(\begin{smallmatrix}
t_0+2u_0\cos2\beta & -t_1-i\sqrt{3}u_2\sin2\beta-u_1\cos2\beta &t_2-u_2\cos2\beta+i\sqrt{3}u_1\sin2\beta \\ 
t_1-i\sqrt{3}u_2\sin2\beta+u_1\cos2\beta & t_{11}+\frac{1}{2}\left(u_{11}+3u_{22}\right)\cos2\beta &-t_{12}+2u_{12}\cos2\beta+i\frac{\sqrt{3}}{2}\left(u_{22}-u_{11}\right)\sin2\beta \\ 
t_2-u_2\cos2\beta-i\sqrt{3}u_1\sin2\beta & t_{12}-2u_{12}\cos2\beta+i\frac{\sqrt{3}}{2}\left(u_{22}-u_{11}\right)\sin2\beta & t_{22}+\frac{1}{2}\left(3u_{11}+u_{22}\right)\cos2\beta
\end{smallmatrix}\right),
\end{eqnarray}
\begin{eqnarray}
h_4=\left(\begin{smallmatrix}
2r_0\cos\beta & i\left(r_1+r_2\right)\sin\beta-\left(r_1-r_2\right)\cos\beta &-\frac{r_1+r_2}{\sqrt{3}}\cos\beta+i\frac{r_1-r_2}{\sqrt{3}}\sin\beta \\ 
i\left(r_1+r_2\right)\sin\beta+\left(r_1-r_2\right)\cos\beta & 2r_{11}\cos\beta &i2r_{12}\sin\beta \\ 
-\frac{r_1+r_2}{\sqrt{3}}\cos\beta-i\frac{r_1-r_2}{\sqrt{3}}\sin\beta & i2r_{12}\sin\beta & \left(2r_{11}+4\frac{r_{12}}{\sqrt{3}}\right)\cos\beta
\end{smallmatrix}\right),
\end{eqnarray}
\begin{eqnarray}
h_5=\left(\begin{smallmatrix}
u_0 & -u_1 & u_2\\ 
u_1 & u_{11} & -u_{12}\\ 
u_2 & u_{12} & u_{22}
\end{smallmatrix}\right).
\end{eqnarray}
In this way the Bogliubov-de Gennes Hamiltonian for the superconducting monolayer NbSe$_2$ ribbon can be obtained as
\begin{eqnarray}
H^{\text{BdG}}_{\text{ribbon}}=\begin{pmatrix}
H_{\text{ribbon}}\left(k_y\right)-\mu I_{6N} &-i\sigma_y\Delta\otimes I_{3N} \\ 
i\sigma_y\Delta\otimes I_{3N} & -H_{\text{ribbon}}^{\ast}\left(-k_y\right)+\mu I_{6N}
\end{pmatrix}.
\end{eqnarray}
With this Hamiltonian, the spectral function $A=\text{Tr}\left[\text{Im}G\left(E, \bm{R}\right)\right]$ for the semi-infinite NbSe$_2$ ribbon with armchair edge is calculated in Fig. 2. Here $\bm{R}$ is the last slab of the NbSe$_2$ ribbon and the Green's function $G=\left(E+i0^{+}-H_{\text{ribbon}}^{\text{BdG}}\right)^{-1}$ can be obtained through the surface Green function method~\cite{Datta}. We set $\Delta=0.02$eV in Fig. 2-4 of the main text.

\begin{figure}
\includegraphics[width=2.6in]{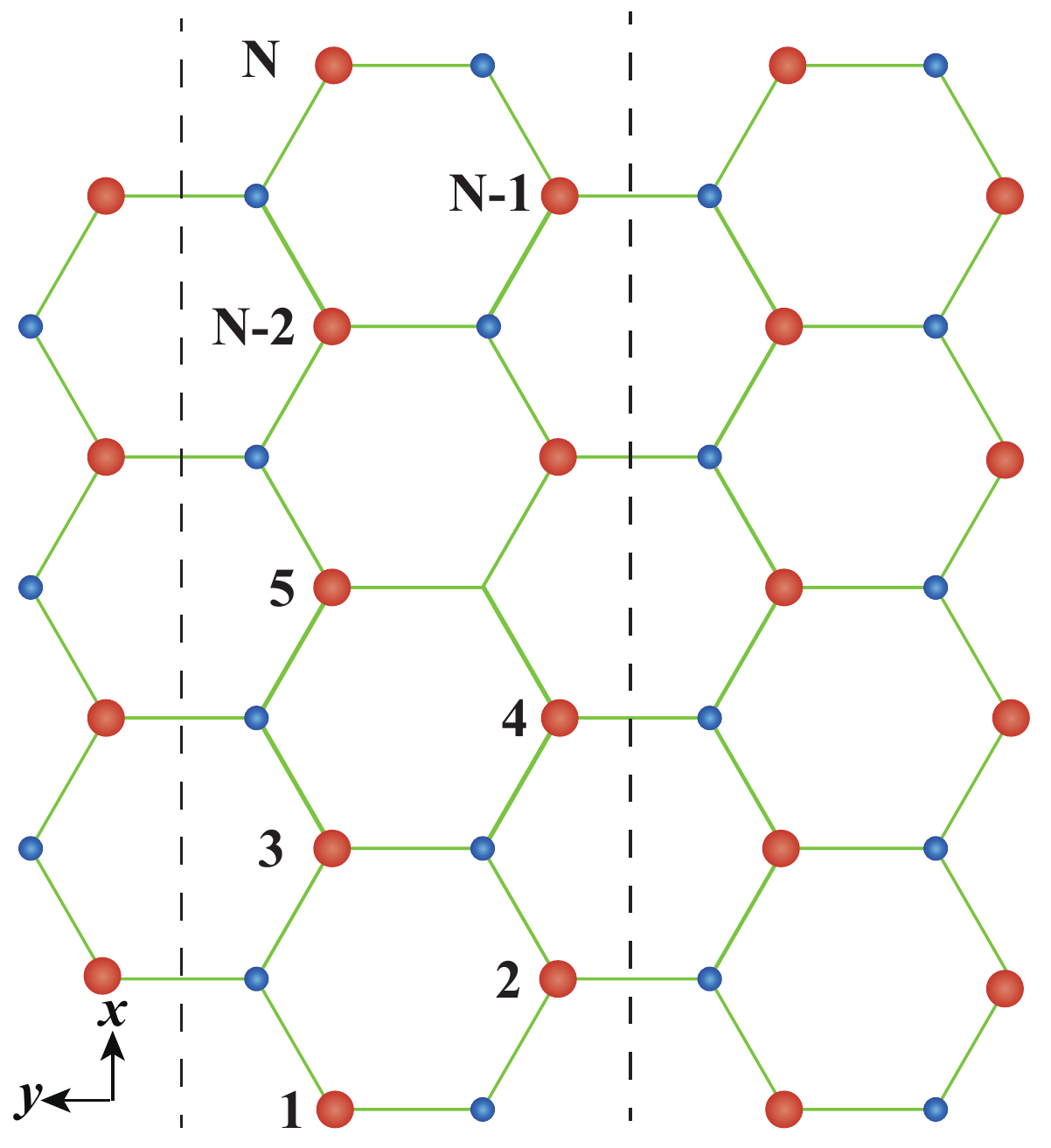}
\caption{The unit cell NbSe$_2$ nano-ribbon with armchair edge.}
\label{FIGS2}
\end{figure}


\begin{thebibliography}{99}





























\bibitem{Ye2} J. M. Lu, O. Zheliuk, I. Leermakers, N. F. Q. Yuan, U. Zeitler, K. T. Law, and J. T. Ye, \emph{Evidence for two-dimensional Ising superconductivity in gated MoS$_2$}, Science {\bf350}, 1353-1357 (2015).

\bibitem{Iwasa2} Y. Saito, Y. Nakamura, M. S. Bahramy, Y. Kohama, J. Ye, Y. Kasahara, Y. Nakagawa, M. Onga, M. Tokunaga, T. Nojima, Y. Yanase, and Y. Iwasa, \emph{Superconductivity protected by spin-valley locking in ion-gated MoS$_2$}, Nat. Phys. {\bf12}, 144-149 (2016).

\bibitem{Mak2} X. Xi, Z. Wang, W. Zhao, J.-H. Park, K. T. Law, H. Berger, L. Forro, J. Shan, and K. F. Mak, \emph{Ising pairing in superconducting NbSe$_2$ atomic layers}, Nat. Phys. {\bf12}, 139-143 (2016).

\bibitem{King} L. Bawden, S. P. Cooil, F. Mazzola, J. M. Riley, L. J. C.-Mclntyre, V. Sunko, K. W. B. Hunvik, M. Leandersson, C. M. Polley, T. Balasubramanian, T. K. Kim, M. Hoesch, J. W. Wells, G. Balakrishnan, M. S. Bahramy and P. D. C. King, \emph{Spin-valley locking in the normal state of a transition-metal dichalcogenide superconductor}, Nat. Commun. {\bf7}, 11711 (2016).

\bibitem{Rashba0} E. I. Rashba, \emph{Symmetry of bands in wurzite-type crystals. 1. Symmetry of bands disregarding spin-orbit interaction}, Sov. Phys. Solid. State {\bf1}, 368 (1959).

\bibitem{Sigrist} P. A. Frigeri, D. F. Agterberg, and M. Sigrist, \emph{Spin susceptibility in superconductors without inversion symmetry}, New J. of Phys. {\bf 6} 115 (2004). 

\bibitem{Pasupathy} A. W. Tsen, B. Hunt, Y. D. Kim, Z. J. Yuan, S. Jia, R. J. Cava, J. Hone, P. Kim, C. R. Dean, and A. N. Pasupathy, \emph{Nature of the quantum metal in two-dimensional crystalline superconductor}, Nat. Phys. {\bf12}, 208-212 (2016).

\bibitem{WangJian} Y. Xing et al. \emph{Ising superconductivity and quantum phase transition in macro-size monolayer NbSe$_2$.} Nano Lett. {\bf 17}, 6802 (2017).

\bibitem{ZhangHao} H. Zhang et al. arXiv:1710.10701 (2017).

\bibitem{WangKang} Q. L. He et al. Science {\bf 357} (2017).

\bibitem{Oppen} Y. Oreg, G. Refael and F. von Oppen, \emph{Helical Liquids and Majorana Bound States in Quantum Wires}, Phys. Rev. Lett. {\bf105}, 177002 (2010).

\bibitem{Sarma1} J. D. Sau, R. M. Lutchyn, S. Tewari and S. Das Sarma, \emph{Generic New Platform for Topological Quantum Computation Using Semiconductor Heterostructures}, Phys. Rev. Lett. {\bf104}, 040502 (2010).

\bibitem{Alicea2} J. Alicea, \emph{Majorana fermions in a tubale semiconductor device}, Phys. Rev. B {\bf81}, 125318 (2010).

\bibitem{Law0} K. T. Law, P. A. Lee and T. K. Ng, \emph{Majorana Fermion Induced Resonant Andreev Reflection}, Phys. Rev. Lett. {\bf103}, 237001 (2009).

\bibitem{Coronado} E. N.-Moratalla, J. Island, S. M.-Valero, E. P.-Cienfuegos, A. C.-Gomez, J. Quereda, G. R.-Bollinger, L. Chirolli, J. A. S.-Guillen, N. Agrait, G. A. Steele, F. Guinea, H. S. J. Zant and E. Coronado, \emph{Enhanced superconductivity in atomically thin TaS$_2$}, Nat. Commun. {\bf7}, 11043 (2016).

\bibitem{Xiaoming} Y. Ma, J. Pan, C. Guo, X. Zhang, L. Wang, T. Hu, G. Mu, F. Huang and X. Xie, \emph{Unusual enhancements of B$_c2$ and T$_c$ in the restacked TaS$_2$ nanosheets}, arXiv: 1712.07763.

\bibitem{Noah0} N. F. Q. Yuan, K. F. Mak, and K. T. Law, \emph{Possible Topological Superconducting Phases of MoS$_2$}, Phys. Rev. Lett. {\bf113}, 097001 (2014).

\bibitem{ABINIT} Here the ABINIT code is used to perform the calculation for the band structure of monolayer NbSe$_2$. See also the web page at http://www.abinit.org.

\bibitem{Yao} G.-B. Liu, W.-Y. Shan, Y. Yao, W. Yao, and D. Xiao, \emph{Three-band tight-binding model for monolayers of group-VIB transition metal dichalcogenides}, Phys. Rev. B {\bf88}, 085433 (2013).

\bibitem{Eriksson} S. Lebegue and O. Eriksson, \emph{Electronic structure of two-dimensional crystal from \emph{ab initio} theory}, Phys. Rev. B {\bf79}, 115409 (2009).

\bibitem{supplementary} The supplementary information for the details of the tight binding model.

\bibitem{Crommie} M. M. Ugeda, et al. \emph{Characterization of collective ground states in single-layer NbSe$_2$}, Nat. Phys. {\bf12} 92-97 (2016).

\bibitem{Mak1} X. Xi, L. Zhao, Z. Wang, H. Berger, L. Forro, J. Shan, and K. F. Mak, \emph{Strongly enhanced charge-density-wave order in monolayer NbSe$_2$}, Nat. Nanotechnol. {\bf10}, 765-769 (2015).

\bibitem{Beernsten} E. Revolinsky, G. A. Spiering, and D. J. Beernsten, \emph{Superconductivity in the niobium-selenium system}, J. Phys. Chem. Solids {\bf26}, 1029-1034 (1965).

\bibitem{Toyota} N. Toyota, et al. \emph{Temperature and angular dependences of upper critical fields for the layer structure superconductor 2H-NbSe$_2$}, J. Low. Temp. Phys. {\bf25}, 485-499 (1976).

\bibitem{Clogston} A. M. Clogston, \emph{Upper Limit for the Critical Field in Hard Superconductors}, Phys. Rev. Lett. {\bf9}, 266 (1962).

\bibitem{Fai} K. F. Mak (Penn State university), private communications.

\bibitem{Hernando} D. Huertas-Hernando, F. Guinea, and A. Brataas, \emph{Spin-orbit coupling in curved graphene, fullerenes, nanotubes, and nanotube caps}, Phys. Rev. B {\bf74}, 155426 (2006).

\bibitem{Schnyder} A. P. Schnyder, S. Ryu, A. Furusaki, and A. W. W. Ludwig, \emph{Classification of topological insulators and superconductors in three spatial dimensions}, Phys. Rev. B {\bf78}, 195125 (2008).

\bibitem{Sau} S. Tewari and J. D. Sau, \emph{Topological Invariants for Spin-Orbit Coupled Superconductor Nanowires}, Phys. Rev. Lett. {\bf109}, 150408 (2012).

\bibitem{Law} C. L. M. Wong, J. Liu, K. T. Law, and P. A. Lee, \emph{Majorana flat bands and unidirectional Majorana edge states in gapless topological superconductors}, Phys. Rev. B {\bf88}, 060504 (R) (2013).

\bibitem{Noah} N. F. Q. Yuan, Y. Lu, J. J. He, and K. T. Law, \emph{Generating giant spin currents using nodal topological superconductors}, Phys. Rev. B {\bf 95}, 195102 (2017). 

\bibitem{Rashba} L. P. Gor'kov, and E. I. Rashba, \emph{Superconducting 2D System with Lifted Spin Degeneracy: Mixed Singlet-Triplet State}, Phys. Rev. Lett. {\bf87}, 037004 (2011).

\bibitem{Zhou} Benjamin. T. Zhou, Noah. F. Q. Yuan, H.-L. Jiang, and K. T. Law, \emph{Ising Superconductivity and Majorana Fermions in Transition Metal Dichalcogenides}, Phys. Rev. B {\bf93}, 180501(R) (2016).

\bibitem{Leggett} A. J. Leggett, \emph{A theoretical description of the new phases of liquid $^3$He}, Rev. Mod. Phys. {\bf47}, 331-414 (1976).

\bibitem{James} J. J. He, T. K. Ng, P. A. Lee, and K. T. Law, \emph{Selective Equal-Spin Andreev Reflections Induced by Majorana Fermions}, Phys. Rev. Lett. {\bf112}, 037001 (2014).

\bibitem{Linder} J. Linder, and J. W. A. Robinson, \emph{Superconducting spintronics}, Nat. Phys. {\bf11}, 307-315 (2015).

\bibitem{Eschrig} M. Eschrig, \emph{Spin-polarized supercurrents for spintronics}, Phys. Today {\bf64}, 43-49 (January 2011).

\bibitem{Yokoyama} M. Sato, Y. Tanaka, K. Yada and T. Yokoyama, \emph{Topology of Andreev bound states with flat dispersion}, Phys. Rev. B {\bf83}, 224511 (2011).

\end{thebibliography}

\begin{thebibliography}{1}

\bibitem{Eriksson} S. Lebegue and O. Eriksson, \emph{Electronic structure of two-dimensional crystals from ab initio theory}, Phys. Rev. B {\bf79}, 115409 (2009).

\bibitem{Yao} G.-B. Liu, W.-Y. Shan, Y. Yao, W. Yao and D. Xiao, \emph{Three-band tight-binding model for monolayers of group-VIB transition metal dichalcogenides}, Phys. Rev. B {\bf88}, 085433 (2013).

\bibitem{Datta} S. Datta, \emph{Quantum Transport} (Cambridge Unniversity Press, Cambridge, New York, 2005).

\end{thebibliography}
\end{document}